\documentstyle[12pt,epsfig]{article}

\newcommand{\mrm}[1]{\mbox{\rm #1}}

\newcommand{\be}{\begin{equation}}
\newcommand{\ee}{\end{equation}}
\newcommand{\nn}{\nonumber}
\newcommand{\bea}{\begin{eqnarray}}
\newcommand{\eea}{\end{eqnarray}}

\newcommand{\rfn}[1]{(\ref{#1})}
\newcommand{\Eq}[1]{Eq.~(\ref{#1})}
\newcommand{\gsim}{\ \rlap{\raise 2pt\hbox{$>$}}{\lower 2pt
\hbox{$\sim$}}\ }
\newcommand{\lsim}{\ \rlap{\raise 2pt\hbox{$<$}}{\lower 2pt
\hbox{$\sim$}}\ }

\newcommand{\D}{\Delta}

\newcommand{\matr}{\left( \begin{array}}
\newcommand{\ematr}{\end{array} \right)}
\newcommand{\g}{\gamma}

\newcommand{\np}[1]{Nucl. Phys. {\bf #1}}
\newcommand{\pl}[1]{Phys. Lett. {\bf #1}}
\newcommand{\pr}[1]{Phys. Rev. {\bf #1}}
\newcommand{\prl}[1]{Phys. Rev. Lett. {\bf #1}}

\newcommand{\ptp}[1]{Prog. Theor. Phys. {\bf #1}}

\makeatletter
\if@twoside
\else \oddsidemargin 00pt \evensidemargin 00pt
\fi
\let\@eqnsel = \hfil

\expandafter\ifx\csname mathrm\endcsname\relax\def\mathrm#1{{\rm #1}}\fi
\@ifundefined{mathrm}{\def\mathrm#1{{\rm #1}}}{\relax}

\makeatother

\begin{document}
\thispagestyle{empty}
\null
\hfill TP-USL/97/15

\hfill FTUV/96-24

\hfill hep-ph/9703215

\vskip 1.5cm

\begin{center}
{\Large \bf      
  Heavy neutrino mixing and  single production at Linear
Collider
\par} \vskip 2.em
{\large		
{\sc J. Gluza$^{1}$, J. Maalampi$^2$, M. Raidal$^{3}$ 
and M. Zra{\l}ek$^{1}$
}  \\[1ex] 
{\it $^1$ Department of Field Theory and Particle Physics,
University of Silesia, Uniwersytecka 4, PL-40-007 Katowice, Poland} \\
{\it $^2$ Department of Physics, Theoretical Physics Division,
University of Helsinki, P.O. Box 9, FIN-00014 Helsinki, Finland} \\
{\it $^3$ Department of Theoreticel Physics, University of Valencia
 and IFIC, 46100 Burjassot, Valencia, Spain}
\\[1ex]
\vskip 0.5em
\par} 
\end{center} \par
\vfil
{\bf Abstract} \par
We study the single production of heavy neutrinos via the processes
$e^-e^+ \rightarrow \nu N$ and $e^-\gamma \rightarrow W^-N $ at future
linear colliders. As a base of our considerations we take a wide class
of models, both with vanishing and non-vanishing left-handed Majorana
neutrino mass matrix $m_L$. We perform a model independent analyses of
the existing experimental data and find connections between the
characteristic of heavy neutrinos (masses, mixings, CP eigenvalues) and
the $m_L$ parameters. We show that with the present experimental
constraints heavy neutrino masses almost up to the collision energy can
be tested in the future experiments.
\par
\vskip 0.5cm
\noindent February 1997 \par
\null
\setcounter{page}{0}
\clearpage

\section{Introduction}

\vspace*{0.5cm}

While the properties of charged fermions are tested with a very high
accuracy the situation in the neutral fermion sector remains poorly
understood. The issues of neutrino masses and mixings are still
unsettled. In the Standard Model (SM) neutrinos are  predicted to be
massless. However, the solar
\cite{solar} and atmospheric \cite{atm} neutrino deficites as well as
the measurement of COBE satellite of the hot component of dark matter
\cite{cobe} seems to indicate that neutrinos do have  small but
non-vanishing masses. All laboratory experiments so far have failed to
measure these masses, having allowed one only to set upper limits  on
their values
\cite{pdb}.   If neutrinos are, indeed, massive there are two
fundamental questions to be answered. Firstly, what is the nature of
their masses i.e. whether neutrinos are Majorana or Dirac particles
and, secondly, why neutrino masses are so tiny compared with the masses
of charged leptons. Both of these questions might show in a new light
if new neutrino species with a large mass were discovered.

 The generic neutrino mass matrix allowed by gauge symmetry is of the
form
\bea M=\matr{cc} m_L & m_D \\
      m_D^T & M_R \ematr ,
\label{matr}
\eea 
where $m_D$ is the submatrix of Dirac type masses and $m_L$ and
$M_R$ are the submatrices of Majorana type masses for left- and
right-handed neutrinos, respectively. The exact form  of the mass
matrix (\ref{matr}) depends on  specific model under consideration. As
suggested by experimental data,  there is a strong hierarchy among
different types of masses. While the Dirac masses forming the matrix
$m_D$ are naturally of the order of charged lepton masses, 
non-observation of right-handed neutrinos forces the mass scale of
$M_R$ to be larger than the mass $M_Z$ of the neutral weak boson
\cite{mz}.  In the SM supplemented with right-handed neutrino states  
the left-handed Majorana masses are zero at tree level (small radiative
corrections of the order  $m_L \sim {\cal O} (1)$ eV can be expected).
However, in  models with left-handed triplet Higgs representations,
e.g. in the left-right symmetric models (LRM) \cite{lrm},
$m_L=h_M v_L,$ where $h_M$ are unknown triplet Yukawa coupling constants
and $v_L$ is the vev of the left-handed triplet field which, due to its
contribution to the  parameter $\rho= M_W^2/(M_Z^2\cos^2\theta_W)$, is
constrained to be below 9 GeV \cite{desp}. The bounds on $h_M$ are not
particularly  restrictive at the moment but will improve considerably
in future  experiments \cite{meie2}. At present we know that $m_L$
cannot exceed ${\cal O}(1)$ GeV.

The physical neutrino states, including the ordinary light neutrinos
$\nu_e$,
$\nu_{\mu}$ and $\nu_{\tau}$, are found by diagonalizing the matrix
(\ref{matr}). Observable variables emerging as a result of the
diagonalization are the masses, mixings and CP parities of the mass
eigenstate neutrinos. Particularly interesting are the mixings between
the light and heavy neutrinos, for which there exist several
constraints from various low-energy measurements, such as the
universality of lepton couplings and the negative searches fo
neutrinoless double beta decay.

In this letter we study a single heavy neutrino production in future
linear colliders (LC) in reactions
\bea e^-e^+ & \rightarrow & \nu N, \label{pr1} \\ e^-\gamma &
\rightarrow & W^- N, \label{pr2}
\eea taking into account the existing constraints on the mixing between
$\nu_e$ and the heavy neutrino $N$,
 assuming both $e^+e^-$ and $e^-\g$ collision options,  and
taking into account polarizations of the initial state particles. The
importance of these reactions stems from the possibility to extend the
kinematical discovery limit of heavy neutrinos almost by a factor of
two when compared with the mass reach of the pair production processes.
In the LC the photon beam can be obtained by scattering intensive laser
pulses off the electron beam \cite{telnov1}. In the case of linearly
polarized laser light the energy spectrum of hard photons is strongly
peaked at
$84\%$ of the electron beam energy and its polarization rate is
essentially  the same as the electron beam one \cite{telnov2}. We will
take into account the polarization of the initial state particles.

 The cross sections of the processes (\ref{pr1}) and (\ref{pr2})  are
proportional to  the light and heavy neutrino mixing angles. In the
see-saw models \cite{seesaw}  these mixings are predicted to be very
small. However, there are other models, e.g. based on $E_6$ and
$SO(10)$ symmetry groups, where the light neutrinos are predicted to be
massless at tree level by symmetry arguments
\cite{nonsee,pr9}, in which case the neutrino mixing angles  are not
related to their masses as  in the see-saw models  and can be
considerably larger allowing for observable effects. In order to cover
all these possible scenarios we perform a model independent
phenomenological study of the allowed space of parameters  describing
heavy neutrinos and relate them to the cross sections of the processes
\rfn{pr1} and \rfn{pr2}.

\section{Experimental bounds on heavy neutrino mixings}

\vspace*{0.5cm}

The recent discussions of the bounds on the mixings between light and
heavy neutrinos have based on simplified analyses as far as the
Majorana mass submatrix $m_L$ is concerned. Either the conditions
ensuing from
$m_L$ have been neglected \cite{pr9} or the entire matrix have been
taken zero \cite{pr7}. In the following analysis we will include the
effects connected to the existence of a non-zero submatrix $m_L$.

Let us denote by $K$ the lepton analogy of the Kobayashi-Maskawa
matrix. The experimental bounds on the elements of that matrix, $K_{\nu
e}$ and $K_{Ne}$, desribing the mixing of $\nu_e$ with light and heavy
neutrinos, respectively, can be summarized as follows:
\bea
\sum\limits_{N(heavy)} \left| K_{Ne} \right|^2 &\leq& \kappa^2=0.0054,
\label{con1} \\
\left| \sum_{\nu(light)}K_{\nu e}^2m_{\nu} \right| &\leq&
\kappa^2_{light}= 0.65\;\mrm{eV},
\label{con2} \\
\left| \sum\limits_{N(heavy)} K_{Ne}^2\frac{1}{m_N} \right| &\leq&
\omega^2 =(2-2.8) \cdot 10^{-5} \;\mrm{TeV}^{-1}.
\label{con3}
\eea 
Here $m_{\nu}$ and $m_N$ denote the masses of light and heavy
neutrinos, respectively.   The first constraint comes from the LEP and
low-energy measurements of lepton universality  \cite{low}, and the
constraints (\ref{con2}) and (\ref{con3}) from the lack of positive
signal in  neutrinoless double beta decay  mediated by the light
\cite{bal} and heavy neutrinos\footnote{To find these values,
complicated analyses of nuclear matrix elements must be performed. It
has been argued in Ref. \cite{mink} that the bounds on $\omega^2$ can
be more than one order of magnitude less restrictive than we present
here. However, our considerations will not be affected by this change.
It is important for our discussion that such a bound can be derived
from the experimantal data \cite{ver}.}
\cite{bb}, respectively. The constraint (4) is valid for both  Dirac
and Majorana  neutrinos, while the conditions (5) and (6), which
follows from neutrinoless double beta decay, give restriction only
for Majorana neutrinos. 

Diagonalization of the matrix (\ref{matr})
yields a relation
\bea
\sum\limits_{\nu (light)}K_{\nu
e}^2m_{\nu}+\sum\limits_{N(heavy)}K_{Ne}^2m_N ={(m_L)}_{\nu_e \nu_e}
\equiv \langle m_L\rangle,
\label{con4}
\eea which together with \Eq{con2} gives the following constraint on
the parameters of heavy neutrinos
\begin{equation}
\left| \langle m_L\rangle- \sum\limits_{N}K_{Ne}^2m_N \right| <
\kappa_{light}^2.
\label{con5}
\end{equation} 
The inequalities \rfn{con1}, \rfn{con3} and \rfn{con5}
establish the parameter space still allowed by the experimental data.
This space depends on the values of $\langle m_L\rangle$, as well as
onthe number and CP properties of the right-handed neutrinos and their
mass spectrum. Therefore, for the heavy neutrino masses $m_N$ testable
at LC one can find the range of allowed values of $\langle m_L\rangle$.
To shorten our notation let us denote the masses of heavy neutrinos by
$M_1\equiv M,$ $M_2=AM,$ and $M_3=BM$, with
$A, B\geq 1$  (we will consider only the cases of at most three heavy
neutrinos),
  and define the new parameters
\be
\D=\frac{\langle m_L\rangle}{M} \;\; , \;\;
\delta=\frac{\kappa^2_{light}}{M}.
\ee We will assume that CP is conserved and define unphysical CP phases
of charged leptons in such a way that the neutrino mixing angles are
purely real
$(K_{N_ie}=x_i)$ if the CP parity of heavy neutrino is
$\eta_{CP}(N_i)=+i$ and purely imaginary
$(K_{N_ie}=ix_i)$ if $\eta_{CP}(N_i)=-i$. In our notation the CP-parity
of the lightest heavy neutrino is always $+i.$ This can be arranged
without loss of generality by a proper redefinition of the CP
eigenstates.

The number $n_R$ of heavy neutrino species varies in different models.
Let us consider the bounds for the $\Delta$ and the $K_{Ne}$ mixing
parameters separately in the cases where this number is $n_R=1$, 2 or 3.

$\bullet\;n_R=1$

From the inequalities \rfn{con1}, \rfn{con3} and \rfn{con5} we obtain
\bea -\delta < \Delta < min(\kappa^2,\omega^2M)+\delta ,
\eea implying that relatively small values of $\langle m_L\rangle ,$
e.g.,
$|\langle m_L\rangle | \leq 2 \cdot 10^{-4}$ GeV for $M=100$ GeV, are
tolerated by data. In this case the largest possible value of
$K_{N_1e}$ is restricted to be
\bea {(K_{N_1e})}_{max}^2=min(\Delta+\delta,\omega^2M,\kappa^2).
\label{n1}
\eea

\newpage 

$\bullet \; n_R=2$

The light-heavy mixing angles are restricted by the inequalities which
depend on the CP eigenvalues of the heavy neutrinos as
\begin{eqnarray} x_1^2+x_2^2 & \leq & \kappa^2,  \nonumber \\
\left| \Delta-x_1^2 \mp Ax_2^2 \right| & \leq & \delta,
\label{ineq2} \\
\left| -x_1^2 \mp \frac{x_2^2}{A} \right| & \leq & \omega^2 M, \nonumber
\end{eqnarray} where the signs $-(+)$ stand for the same (different)
values of $\eta_{CP}$ of the neutrinos. From these inequalities we
obtain bounds on acceptable values of
$\Delta$ and $x_1^2$ as follows.

(i) $\;\;$ If $\eta_{CP}$'s of both neutrinos are the same then
\bea -\delta \leq \Delta \leq min(A\kappa^2,A^2\omega^2M)+\delta ,
\eea and the largest allowed value of $K_{N_1e}$ can be found from the
expression
\bea {(x_1^2)}_{max}=min \left( \Delta+\delta,
\frac{A^2\omega^2M-\Delta+\delta}{A^2-1},
\frac{A\kappa^2-\Delta+\delta}{A-1} \right) < min(\omega^2 M, \kappa^2).
\label{n2i}
\eea

(ii) $\;\;$ If $\eta_{CP}(N_1)=-\eta_{CP}(N_2)=+i$ the allowed values
of $\Delta$ are in the range
\bea -min(A\kappa^2,(A-1)\kappa^2+A\omega^2M)-\delta \leq \Delta \leq
min(\omega^2M, \kappa^2) + \delta ,
\eea and the maximal value of $K_{N_1e}$ is given by
\begin{equation} {(x_1^2)}_{max}=min \left[
\frac{A\kappa^2+\Delta-\delta}{A+1},
\frac{-\Delta+A^2\omega^2 M+\delta}{A^2-1} \right] \leq min \left[
\frac{\kappa^2+A\omega^2M}{A+1},\kappa^2 \right] .
\end{equation}

$\bullet \; n_R=3$

In this case the inequalities take a form
\begin{eqnarray} x_1^2+x_2^2+x_3^2 & \leq & \kappa^2 , \nonumber \\
\left| \Delta-x_1^2 \mp Ax_2^2 \mp Bx_3^2 \right| & \leq & \delta
\label{ineq3} \\
\left| -x_1^2 \mp \frac{x_2^2}{A}
\mp \frac{x_3^2}{B} \right| & \leq & \omega^2 M. \nn
\end{eqnarray} As $\Delta$ can be positive or negative  and $A>B$ or $A
\leq B,$ there are only three independent CP configurations: CP
eigenvalues of all neutrinos are the same,
$\eta_{CP}(N_1)=\eta_{CP}(N_2)=-\eta_{CP}(N_3)$ or
$\eta_{CP}(N_1)=-\eta_{CP}(N_2)=-\eta_{CP}(N_3)$.

(i) $\;\;$ The case of  equal CP eigenvalues of all heavy neutrinos is
qualitatively the same as for $n_R=2$ and we obtain
\bea -\delta < \Delta \leq max \left\{
min(A\kappa^2,A^2\omega^2M),min(B\kappa^2, B^2\omega^2M) \right\} +
\delta ,
\eea and
\begin{eqnarray} {(x_1^2)}_{max} & = & min  \left\{ \Delta+\delta, max 
\left[ min \left(
\frac{B^2\omega^2M-\Delta+\delta}{B^2-1},\frac{B\kappa^2-\Delta+\delta}{B-1}
\right), \right. \right. \nonumber \\ && \hspace{1 cm} min \left. \left.
\left( \frac{A^2\omega^2M-\Delta+\delta}{A^2-1},
 \frac{A\kappa^2-\Delta+\delta}{A-1} \right)
\right] \right\} \leq min(\omega^2 M, \kappa^2).
\label{n3i}
\end{eqnarray}

(ii) $\;\;$ The allowed region of the neutrino mixing angles in the
case $\eta_{CP}(N_1)=\eta_{CP}(N_2)=-\eta_{CP}(N_3)=i$ is sketched in
Fig. 1. The solutions belong to
$\Omega_2$ plane which lay within $\Omega_1$ and $\Omega_3$ planes. Its
situation depends on the value of $\D.$ The constraints on $\D$ are
different if $A>B$ or $A \leq B$. For $A>B$ we obtain
\bea -min\left\{ B\kappa^2, max \{ B^2\omega^2M,
(B-1)\kappa^2+B\omega^2M \}
\right\} - \delta  \leq \nonumber \\
\leq \Delta
\leq min\{ A\kappa^2,(A-B)\kappa^2+AB\omega^2M \}+\delta ,
\label{n3iiad}
\eea and for $A \leq B$
\bea -min\left\{ B\kappa^2, (B-1)\kappa^2+B\omega^2M \right\} -\delta
\leq \Delta
\leq min\{ A\kappa^2, A^2\omega^2M \}+\delta.
\label{n3iibd}
\eea Independently of the relation between $A$ and $B$ the largest
possible value of $K_{N_1e}^2$ is given by
\begin{eqnarray} &&{(x_1^2)}_{max}  = min  \left[
\frac{B\kappa^2+\Delta-\delta}{1+B},max \left[
\frac{B^2M\omega^2-\Delta+
\delta}{B^2-1},
\frac{(A-B)\kappa^2+ABM\omega^2-\Delta+\delta}{(A-1)(B+1)}
                              \right] \right] \nonumber \\ && \leq min
\left[
          \frac{\kappa^2+BM\omega^2}{B+1}, \kappa^2 \right] .
\label{n3ii}
\end{eqnarray}

(iii) $\;\;$ The case $\eta_{CP}(N_1)=-\eta_{CP}(N_2)=-\eta_{CP}(N_3)=i$
is similar to $n_R=2$ (ii) one. The region of allowed values of
$\Delta$ is given by
\begin{eqnarray} &&-max\left\{ min [ A\kappa^2,
(A-1)\kappa^2+A\omega^2M ], min[ B\kappa^2, (B-1)\kappa^2+B\omega^2M ]
\right\} - \delta \leq \nonumber \\ &&\hspace{6 cm} \leq \Delta \leq min
(\omega^2M, \kappa^2)+\delta ,
\label{n3iiid}
\end{eqnarray} and the maximally allowed  mixing $K_{N_1e}^2$ can be
found from
\begin{eqnarray} &&{(x_1^2)}_{max}  = max  \left\{ min \left[
\frac{A^2\omega^2M-\Delta+\delta}{A^2-1},\frac{A\kappa^2+\Delta-\delta}{A+1}
\right], \right. \\ && \;\;\;\;\;\;\; min \left.  \left[
\frac{B^2\omega^2M-\Delta+\delta}{B^2-1},
        \frac{B\kappa^2+\Delta-\delta}{B+1} \right]
\right\} \leq max  \left\{
\frac{\kappa^2+A\omega^2M}{A+1},\frac{\kappa^2+B\omega^2M}{B+1}
\right\}. \nonumber
\label{n3iii}
\end{eqnarray} All the cases considered above showed that the regions
of allowed values of
$\D$ depend on the number of heavy neutrinos, their CP properties and
masses in a rather complicated way\footnote{Of course, one could follow
a different approach: take some theoretical model for $m_L$ as a
starting point and then use the experimental data to constrain the
possible spectrum of heavy neutrinos and their CP parities.}. In
general,
$\D$ can be both positive and negative and in magnitude as large as
several times of
$\kappa^2$ depending on the values of $A$ and $B.$ Letting $\D,$
$A,$ $B$ and neutrino masses to vary one can obtain large neutrino
mixing angles which provide observable effects in the LC experiments.

\section{Single heavy neutrino production at LC}

\vspace*{0.5cm}

The processes \rfn{pr1}, \rfn{pr2} occur entirely due to the
light-heavy neutrino mixings and, therefore, their cross sections are
proportional to the mixing factors $K_{N_ie}$. The situation is
simplest when there exists just  one heavy neutrino with which the
elctron neutrino mixes, i.e.
$n_R=1$, or when all the heavy neutrinos have the same CP parities
since the experimental bound \rfn{con3} constraines $K_{N_1e}^2$ to be
very small (see Eqs. \rfn{n1}, \rfn{n2i}, \rfn{n3i}). In these cases 
the cross sections are very small, e.g., for the process
$e^-e^+ \rightarrow \nu N$ 
$\sigma=0.8(1.0)\;$fb ($M_N=500$ GeV and $\sqrt{s}=1(2)$ TeV) and with
the anticipated integrated luminosities of L=10-100 fb$^{-1}$
\cite{snow} it will be practically impossible to detect these
processes. Therefore, we have to study the cases of several heavy
neutrinos with differing CP eigenvalues.

In order to obtain large mixing angle ${(x_1^2)}_{max}$  and to
illustrate its dependence on $\Delta$ we shall consider the most
natural situation $n_R=3$ for two cases
$\D= 0$ and $\D\neq 0.$  In the latter case the appropriate range of
$|\D|$ can be obtained by choosing, e.g.,
$|\langle m_L\rangle |={\cal O}(1)$ GeV and varying heavy neutrino mass
in the range testable at the LC, $M=0.1-1$ TeV.  Let us first study the
case (ii) with $A>B.$ If $\D=0$ then it follows from \Eq{n3ii} that for
quite large range of $A$ and $B$ mixing angles close to the absolut
maximum are allowed. In particular, if $B=1$ (i.e., two Majorana
neutrinos are degenerate and form one Dirac neutrino) then the mixing
angle can be as large as
$\kappa^2/2=0.0027$ independently of the value of $A.$ If we allow for
non-zero $\D$  according to
\Eq{n3iiad} we see that its large positive values yield small mixing
angles. However, coosing $\D=-(B-1)\kappa^2$ we obtain in the limit
$\delta \ll \kappa^2$
\bea {(K_{N_1e}^2)}_{max}=\frac{\kappa^2}{B+1}.
\label{maxmix}
\eea As we see, in this case there is a continous range of  relatively
large  possible mixing angles $K_{N_1e}.$

If $B>A$ the situation is somewhat different. According to \Eq{n3ii},
if $\D=0$ the only possibility to have sizable mixing angles is to have
$B=1$ which yields $(x^2_1)_{max}=\kappa^2/2.$ If $B\neq 1$ the maximum
mixing angle decreases very quickly to a value proportional to
$\omega^2$. However, for $\D\neq 0$ one can have a continous range of
large mixing angles, described by \Eq{maxmix} for the minimal $\D$. The
results for the case  (iii) are very similar to the latter ones (here
$A$ and $B$ are symmetric).

Note that the experimental bounds \rfn{con1}-\rfn{con3} correlate the
allowed values of $\D$ with neutrino masses and mixings. While for the
previously given representative values of $\langle m_L\rangle $ and $M$
one can have any value of $A$ and $B$ then for the larger $\langle
m_L\rangle $ which give, e.g.,  $|\D|=0.1$ small values of $A,B$ are
excluded. Consequently, in this case the maximally allowed mixing
angles are suppressed by large $A,B.$ However, since $M$ is expected to
be large, it would be difficult accommodate so large values of $\D$
with the existing phenomenology.

The discussion above can be summarized as shown in Fig. 2 where we plot
properly normalized $(x^2_1)_{max}$ against $B.$ The maximal mixing
angles for $\D=0$ in the cases (ii) $B>A$ and (iii), which can be close
to maximum only if $B\approx 1,$ are described by short-dashed line.
For the other cases, i.e. $\Delta\neq 0$ or case (ii)
$A>B$, one has a variety of possibilities to obtain mixings close to the
global maximum which is presented in figure by solid line. In
particular, an extreme situation with $\D=-0.1$ is depicted with
long-dashed line.

To investigate the viability of a single heavy neutrino production at LC
we present the cross sections of the processes $e^+e^- \rightarrow \nu
N$ and $e^-\g\rightarrow  W^-N$ in Fig. 3 and Fig. 4, respectively, for
various collision energies of the LC. For the mixing angle
$|K_{N_1e}|^2$ we choose the  maximal  allowed value of
$\kappa^2/2=0.0027.$ Since the cross sections are proportional to $|
K_{N_1e}|^2 $ one can easily scale their values by an appropriate factor
if the experimantal constraints become stronger. As can be seen in the
figures, the typical cross sections are of the order of 100 fb which
with the anticipated luminosities would mean a few thousand events per
year. Therefore, studies of the processes \rfn{pr1} and \rfn{pr2} would
allow to extend the heavy neutrino mass range testable in the LC
experiments almost up to the kinematical threshold.

\section{Conclusions}

\vspace*{0.5cm}

We have studied the viability of a single heavy neutrino production at
LC via the processes $e^-e^+ \rightarrow \nu N$ and
$e^-\gamma \rightarrow  W^-N.$ Unlike in the previous works we have
taken into account effects of non-vanishing left-handed Majorana
neutrino mass matrix $m_L.$ The result depends in a crucial way on the
number of heavy neutrinos, their mass spectrum, CP eigenvalues and the
value of the left-handed mass matrix parameter $\langle m_L\rangle.$ If
there is only one heavy neutrino or if CP eigenvalues of all heavy
neutrinos are the same then the constraints coming from the negative
search of neutrinoless double beta decay constrain the cross sections
much below the observable limit. In the other cases the neutrino mixing
angle can be large, up to the maximally allowed $\kappa^2/2.$ In
particular, if $\langle m_L\rangle \neq 0$ there is quite large
continous parameter space that allows for observable effects in the LC
experiments. With the present experimental constraints on neutrino
mixings heavy neutrinos masses almost up to the kinematical threshold
will be testable in future colliders.

\subsection*{Acknowledgement}

We thank  A. Santamar\'{\i}a and R. Vuopionper\"a for clarifying
discussions. M.R. thanks the Spanish Ministry of Science and Education
for a postdoctoral grant at the University of Valencia. This work is
supported by CICYT under grant AEN-96-1718 and for two of us (J.G. and
M.Z.) in part by  Polish Committee for Scientific Research, Grant Nos.
PB2P30225206/93 and PB659/P03/95/08. J.G. would like to thank to
the Foundation for Polish Science for the fellowship, too.

\newpage

{\large\bf Figure captions}

\vspace{0.5cm}

\begin{itemize}

\item[{\bf Fig.1.}]  Sketch of allowed mixing angles 
$x_1^2,$ $x_2^2,$ $x_3^2$ for the case $n_R=3$ with 
$\eta_{CP}(N_1)=-\eta_{CP}(N_2)=-\eta_{CP}(N_3)$. Solutions belong to
the
$\Omega_2$ plane which is situated between $\Omega_1$ and $\Omega_3$
ones. Maximal $x_1^2$ is realized when the most protrude point S of the
$\Omega_2$ plane approaches  the  point S'. 
\\

\item[{\bf Fig.2.}]  Dependence of the maximal mixing angle $K_{N_1e}^2$
which is normalized to $\kappa^2/2$ on $B.$  Solid line represents the
global maximum as a  function of $B.$ Short-dashed line describes the
behaviour of maximal mixing angle in the case
$\eta_{CP}(N_1)=-\eta_{CP}(N_2)=-\eta_{CP}(N_3),$ $\D=0$  while
long-dashed line shows a case with very large $\D.$ 
\\

\item[{\bf Fig.3.}]  Maximally allowed cross sections for the process
$e^-e^+\rightarrow \nu N$ for different CM energies as functions of 
the lightest heavy neutrino mass. The mixing angle is taken to be
$K_{N_1e}^2=\kappa^2/2$=0.0027
\\

\item[{\bf Fig.4.}]  Maximally allowed cross sections for the process
$e^-\g\rightarrow NW^-$ for different CM energies as functions of 
heavy neutrino masses. Electron beam is assumed to be  left-handedly
and photon beam $\tau=-1$ linearly polarized.  Curves denoted by $a$
and $b$ present cross sections of the  lightest heavy neutrino
production for $B=1$ and $B=5,$ respectively.

\end{itemize}

\newpage

\begin{figure}[b]
\begin{center}
\mbox{\epsfxsize=16cm\epsfysize=12cm\epsffile{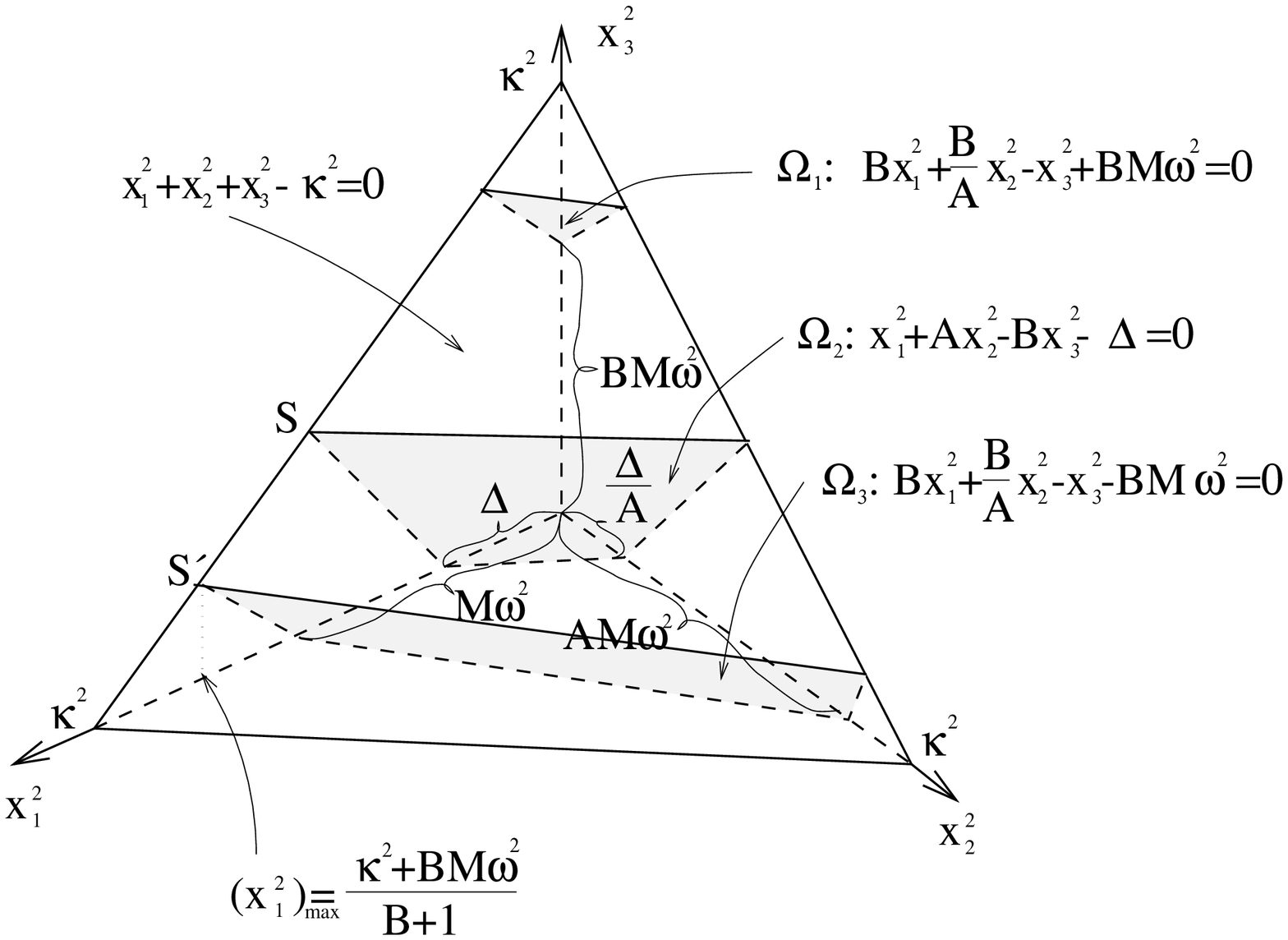}}
\caption{ }
\end{center}
\end{figure}

\newpage

\begin{figure}[ht]
\begin{center}
\mbox{\epsfxsize=16cm\epsfysize=16cm\epsffile{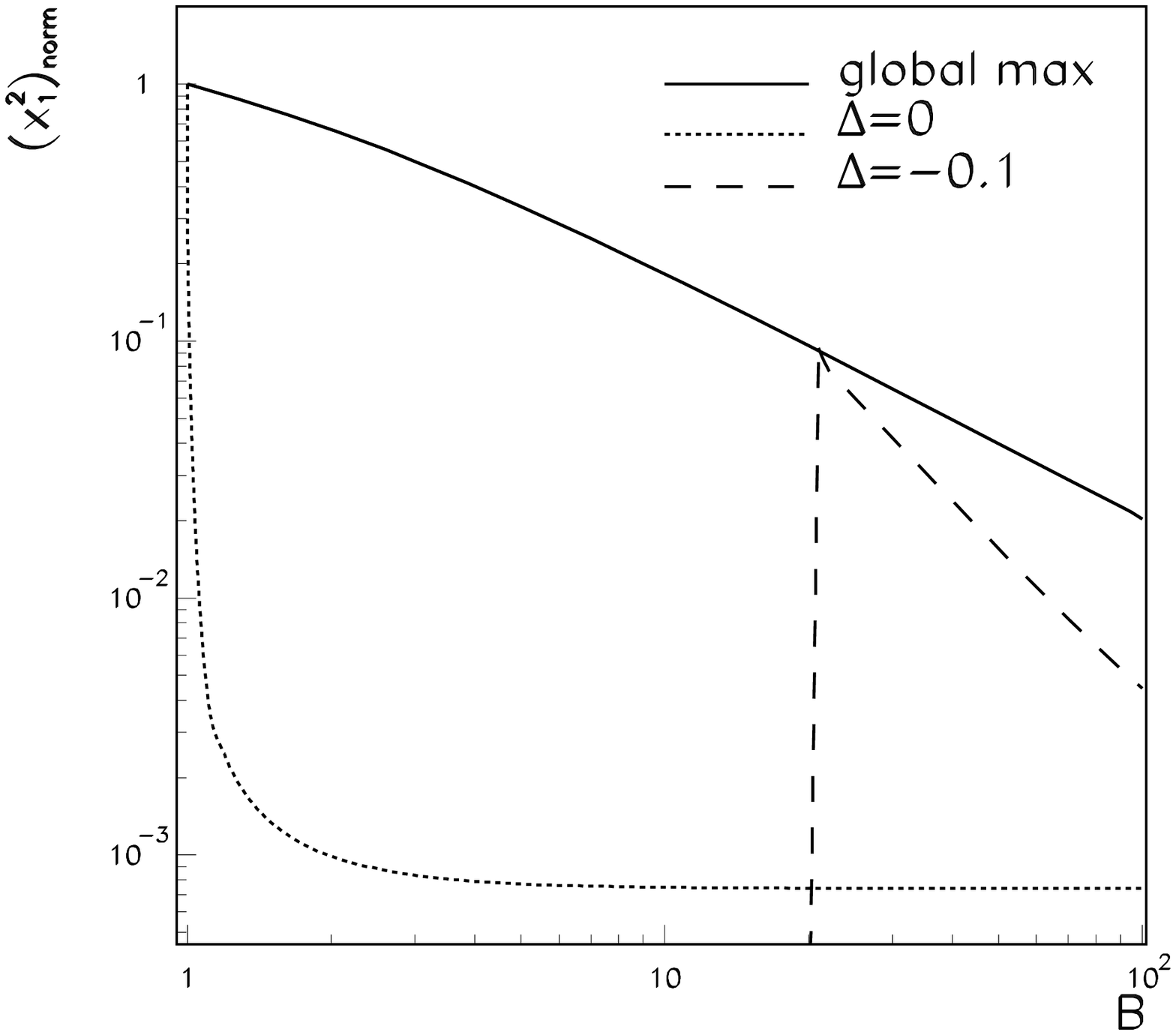}}
\caption{  }
\end{center}
\end{figure}

\newpage

\begin{figure}[ht]
\begin{center}
\mbox{\epsfxsize=16cm\epsfysize=16cm\epsffile{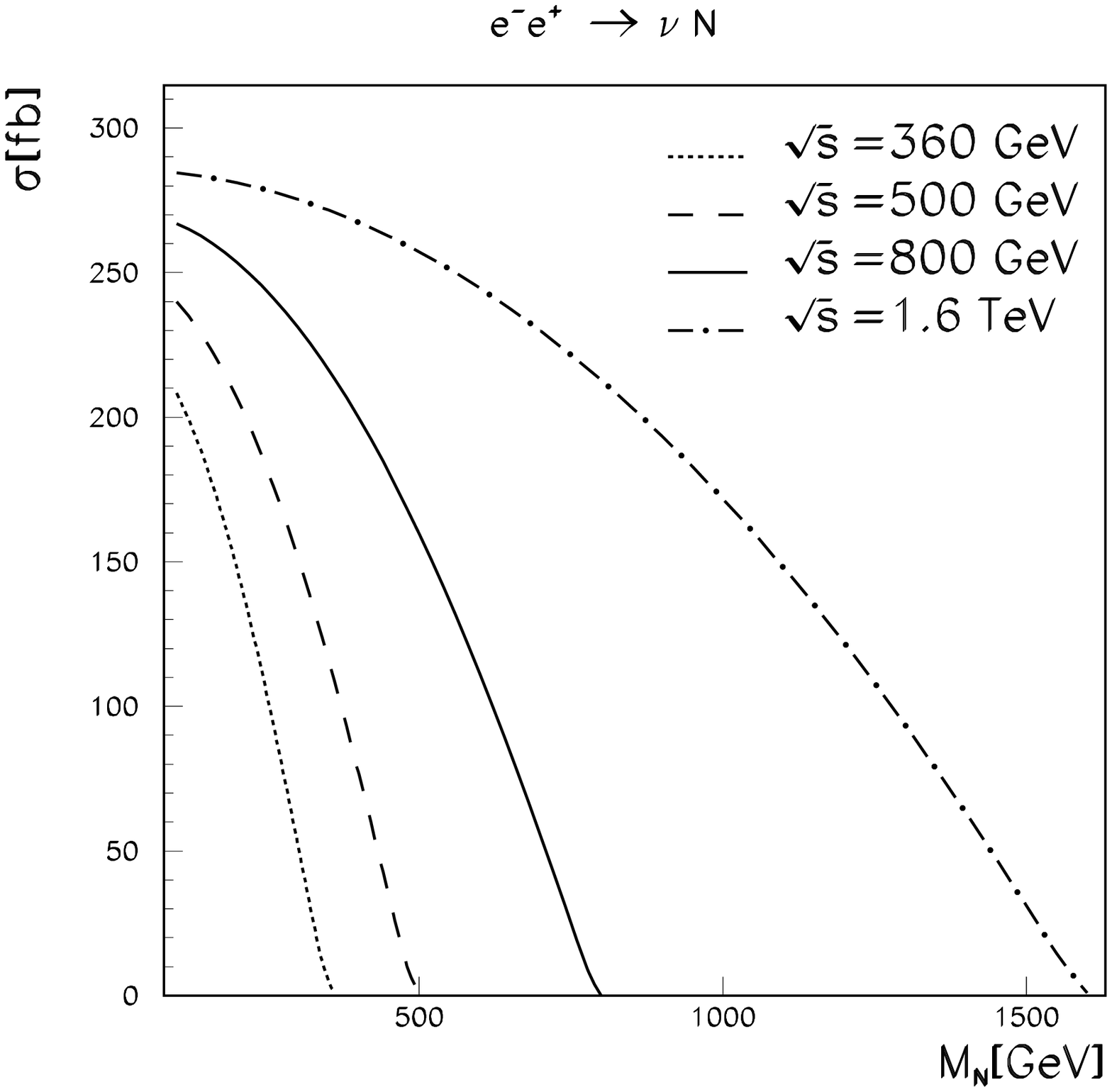}}
\caption{ }
\end{center}
\end{figure}

\begin{figure}[ht]
\begin{center}
\mbox{\epsfxsize=16cm\epsfysize=16cm\epsffile{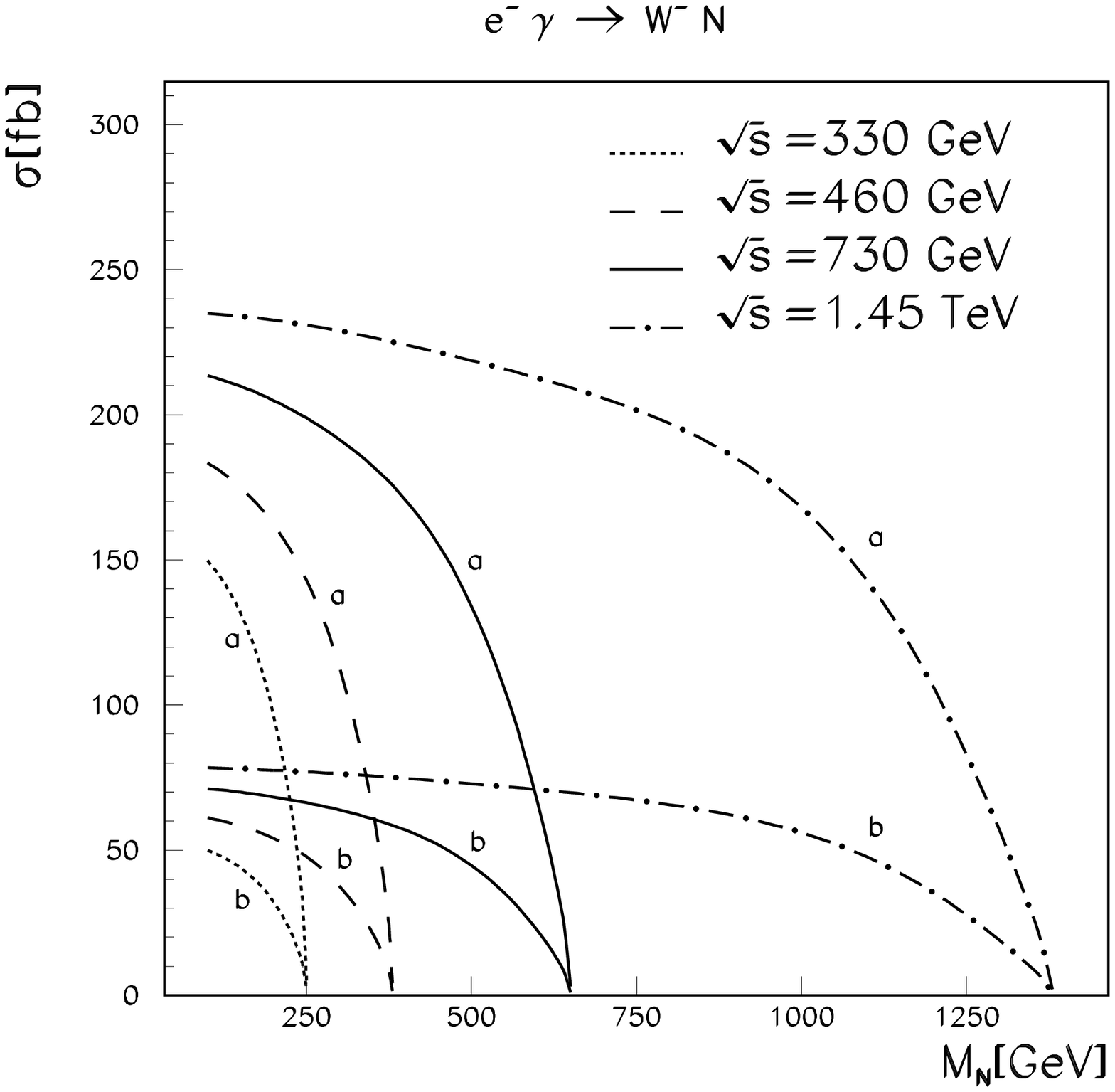}}
\caption{  }
\end{center}
\end{figure}

\end{document}